\documentclass[prl,aps,twocolumn,superscriptaddress,amssymb,amsmath,floatfix,showpacs,notitlepage]{revtex4-1}

\usepackage[dvips]{epsfig}
\usepackage[usenames]{xcolor}
\usepackage{graphicx}
\usepackage{dcolumn}

\newcommand{\be}{\begin{equation}}
\newcommand{\ee}{\end{equation}}
\newcommand{\bea}{\begin{eqnarray}}
\newcommand{\eea}{\end{eqnarray}}
\def\no{\nonumber\\}

\begin{document}

\title{Patterns of polar domains in a spatiotemporal model of interacting polarities}

\author{Maryam Aliee}
\email{maryam.aliee@gmail.com}
\affiliation{Reproduction et D\'eveloppement des Plantes, Universit\'e de Lyon, ENS de Lyon, UCB Lyon 1, INRA, CNRS, 69364 Lyon CEDEX 07, France}
\affiliation{Inria Virtual Plants, CIRAD, INRA, Universite de Montpellier, France}
\affiliation{PULS Group, Institut f\"ur Theoretische Physik and the Excellence Cluster EAM, FAU Erlangen-N\"urnberg, N\"agelsbachstrasse 49b, 91052 Erlangen, Germany}
\author{Arezki Boudaoud}
\affiliation{Reproduction et D\'eveloppement des Plantes, Universit\'e de Lyon, ENS de Lyon, UCB Lyon 1, INRA, CNRS, 69364 Lyon CEDEX 07, France}
\email{Arezki.Boudaoud@ens-lyon.fr}

\date{\today}

\begin{abstract}
Polarity fields are known to exhibit long distance patterns, in both physical and biological systems. 
The mechanisms behind such patterns are poorly understood. 
Here, we describe the dynamics of polarity fields using an original physical model that 
generalizes classical spin models on a lattice by incorporating effective transport of polarity molecules between neighboring sites. 
We account for an external field and for ferromagnetic interactions between sites
and prescribe the time-evolution of the system using two distinct dissipative classes for non-conserved and conserved variables representing polarity orientation and magnitude, respectively. 
We observe two main types of steady-state configurations \--- disordered configurations and patterns of highly polar spots surrounded by regions with low polarity \---
and we characterise patterns and transitions between configurations.
Our results may provide alternative pattern-generating mechanisms for materials endowed with polarity fields.
\end{abstract}

\maketitle

%
It is important to investigate the mechanisms underlying the formation of patterns in polarity fields.
Experiments and models revealed pattern formation in both passive and active media involving polymers or rods~\cite{PhysRevLett.97.090602,PhysRevLett.101.218303,B923942A,Saintillan2008}.
Biological tissues may also be considered as polar materials formed of cells, each expressing distinct polarities~\cite{Thompson:1942,RevModPhys.85.1143,Julicher20073}.
For instance, the asymmetric distribution of a specific protein within a cell or at its periphery defines a polarity~\cite{Simons2008, Okada01071991}.
Planar cell polarity (PCP) is observed in thin tissues and entails the polarization of cells tangentially to the tissue sheet~\cite{EATON2011747}.
PCP often exhibits long distance patterns that play important roles in tissue morphogenesis by regulating cell division, cell flow, and tissue mechanics~\cite{Wang647,Blankenship2006459,HAMANT2010454,Salbreux2012}. 
PIN1 (PIN FORMED1) proteins that enable the transport of a plant hormone, auxin, and PCP complexes in the Drosophila wing are  well-studied examples of polarity fields~\cite{Yang:2008aa, Asnacios2012584,Heisler2010,Zallen20071051,EATON2011747}.
Experimental observations suggest that polarity can be coordinated over a tissue. 
This could occur in two ways. 
All cells could have their own target polarity, each coupled to an external field that prescribes the overall pattern.
To reach a global polarity pattern, this would require a well-defined external field, that may be a fluid flow, gradient of chemical signal, or shear stress~\cite{HARUMOTO2010389,Aigouy2010773,Mani20420}.  
Alternatively, the interactions between the polarities of neighboring cells regulate pattern formation. 
Such cell-cell couplings may operate directly, through membrane-spanning complexes, or indirectly, through diffusing molecules~\cite{Heisler2010,Sauret2013,Zallen20071051,Adler:2002aa,Burak2009,Amonlirdviman423,REEVES2006289,Abley2061}.
For instance, auxin couples the polarity of PIN1 proteins in neigboring cells: PIN1 polarity is enhanced by auxin, which is, in turn, transported by PIN1 proteins.

An important class of biophysical models that describe coupling and patterning of polarities is based on classical physical models of interacting magnetic dipoles.
In particular, Ising, XY, and liquid crystal models have been used to study formation of such patterns~\cite{PhysRevLett.43.744,deGennes:1993}.
All these models successfully explain the alignment between different fields or between a field and an external flow~\cite{Burak2009,Aigouy2010773,Salbreux2012,Abley2061,Mani20420,HARUMOTO2010389}. 
However, the formation of non-uniform or intermittent patterns is still poorly understood. 
In this letter, we extend the classical dipole models and allow for transport of polarity molecules between neighboring cells. 
In a simplified approach, we consider transport of polarity as a proxy for molecular processes that couple the polarities of neighboring cells and transport of molecules interacting with them. 
Including such a possibility in our theoretical model is motivated by interaction of PIN1 polarity and the long-range signal auxin.
We find that transport generates behaviors that differ significantly from those of the classical models~\cite{PhysRevLett.43.744,deGennes:1993,Aigouy2010773}. In particular, our model yields spatial patterns with localized polarity. 

Our model describes a field of polar cells in a two dimensional tissue, assumed to form a square lattice of identical cells. 
In 2D, a polarity can be a vector or a nematic described by an angle, $\phi$, and a magnitude, $\rho$.
These variables are given for individual cells indexed by $\alpha$, $\phi=\phi_\alpha$ and $\rho=\rho_\alpha$.
We formulate the model for nematic polarity fields, though it can easily be extended to vector polarities.  
The simplest interaction energy is given by the inner products of polarities,
\bea
E= -K  \sum_{\langle \alpha, \beta  \rangle } \rho_\alpha \rho_\beta \cos 2(\phi_\alpha-\phi_\beta) 
- F \sum_{ \alpha} \rho_\alpha \cos 2 \phi_ \alpha . \; \; \; \; \;\; \; \; 
\label{eqEn1}
\eea
The first term describes the interaction between neighbors, assuming a positive coupling constant, $K$; the sum is over all pairs of neighboring sites $ \langle \alpha, \beta  \rangle$.
This term is minimized when neighboring polarity nematics align and increase in amplitude. 
The second term corresponds to alignment with an external field $F$ that is parallel to the $x$ axis and the sum is over all sites.
The factor of $2$ in the argument of cosines reflects the nematic nature of polarities and should be removed for vector polarities.

Concerning dynamics, we assume different dissipative properties for orientation and magnitude.
The nematic orientation, $\phi$, follows a stochastic differential equation for a non-conserved variable (model A in~\cite{Lubensky}),
\bea
\frac{d\phi_ \alpha}{dt}=-\xi_a \frac {\delta E}{\delta \phi_ \alpha} +\eta_\alpha (t).
\label{eqDphi}
\eea
Here $\xi_a$ is the dissipative coefficient. $\delta$ depicts the functional derivative, which can be replaced by a partial derivative in a discrete model.
$\eta_\alpha (t)$ is an uncorrelated random noise with zero mean and white noise spectrum.
In the numerics, $\eta$ is a random variable with Gaussian normal distribution and standard deviation $\zeta$, proportional to the square root of an effective temperature (see Supplementary Note). 

Transport of polarity is considered as a proxy of molecular processes involved in site-site or site-external field couplings. Accordingly, we describe the dynamics of polarity magnitude, $\rho$ by a partial differential equation for a conserved variable (Model B in~\cite{Lubensky}),
\bea
\frac{d\rho_ \alpha}{dt}= \xi_\rho \Delta\frac {\delta E}{\delta \rho}|_ \alpha + D \Delta \rho |_ \alpha \; .
\label{eqDd}
\eea
Here $\xi_\rho$ is the corresponding dissipative coefficient and $\Delta$ is the 2D Laplacian operator.
Intercellular transport is the main novelty of our model and will appear to drastically influence the dynamics of the system. 
Transport is not intrinsically active since it is driven by fluxes to minimize the energy function. 
We also account for pure diffusion, with a coefficient $D=\mu k_B T$
that depends on the effective temperature $T$, $\mu$ being a mobility.
This effectively incorporates noise into this equation, without explicitly including stochastic noise as in Eq.~(\ref{eqDphi}).
In this formulation, the total polarity magnitude is conserved although this is not a key assumption in our model. 

The dynamic equations for $\phi$ and $\rho$ can be solved numerically on a lattice of $N \times N$ square cells with periodic boundary conditions (see Supplementary Note). 
The initial conditions for polarity magnitude and orientation in each cell are given by random numbers with uniform distributions.
In the Supplementary Note, we show how the solutions reach quasi-stationary states, with total energy decreasing and reaching a plateau.
We here study states for different choices of parameters.
We present the results as a function of normalized model parameters, though we use the same notations for convenience (see Supplementary Note).
In the following, we analyze separately how each of the two energy terms (Eq.~\ref{eqEn1}) influences the dynamics of the polarity network.

{\it Nonuniform patterns with an external field. }
\begin{figure}
\centerline{\includegraphics[width=7.5cm,angle=0]{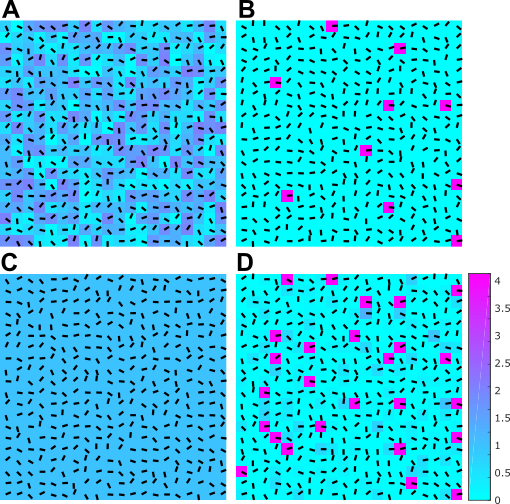}}
\caption{Steady state configurations of the polarity field in a $20\times20$ square lattice  in the presence of an external field. The polarity orientation at each site is shown a black bar. The magnitude of polarity is color-coded with the same scale for all panels. The noise level is $\zeta=1.8$.
Each panel corresponds to different transport properties: (A)  $\xi_\rho=0$, $D=0$, (B)  $\xi_\rho=2$, $D=0$, (C)  $\xi_\rho=0$, $D=0.033$, (D)  $\xi_\rho=2$, $D=0.033$. 
}
\label{figE1}
\end{figure}
We first study the response to an external field with no coupling between neighbors ($K=0$). 
We rewrite Eqs.~(\ref{eqDphi}-\ref{eqDd}) as 
\bea
\frac{d\phi_\alpha}{dt}&=& -2\xi_a F \rho_\alpha \sin 2 \phi_\alpha+\eta_\alpha (t),
\no
\frac{d \rho_\alpha}{dt}&=& -\xi_\rho \Delta f |_{\alpha}+ D \Delta \rho |_ \alpha \;,    \;\;\;\; f_{\alpha}= F \cos 2\phi_{\alpha} \;,
\label{eqD1}
\eea
We first consider an intermediate value of noise
(see hereafter for a more rigorous definition of \textit{intermediate}).
No order can be seen for static polarity magnitude when $\xi_\rho=0$, without ($D=0$, Fig.~\ref{figE1}A) or with diffusion between neighbors ($D>0$, Fig.~\ref{figE1}C), when
polarity magnitude is homogenous and polarity nematic is randomly oriented.
In contrast, the transport of polarity molecules between neighbors ($\xi_\rho>0$) leads to a configuration with inhomogeneous distributions of polarity (Fig.~\ref{figE1}B,D). 
Starting from a random initial state, polarity magnitude increases in cells that are initially better aligned with the external field, due to intercellular transport.  
In parallel,  due to lower interaction energy in cells with larger polarity magnitude, fluctuations of orientation are reduced, enhancing alignment with the external field and amplifying polarity.
This feedback loop yields domains of high polarity magnitude, in which the orientation follows the external field,
while elsewhere, polarity magnitude remains low and orientation appears random.
\begin{figure}
\centerline{\includegraphics[width=8cm,angle=0]{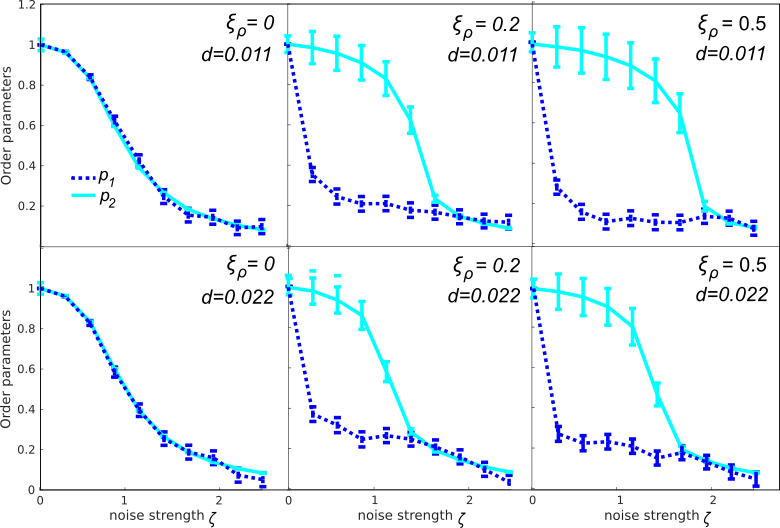}}
\caption{Influence of model parameters on order parameters ($p_1$ and $p_2$). We assume the diffusion constant proportional to the square of noise strength, $D=d\zeta^2$. Error bars represent the standard error of the mean (SEM).}
\label{figEO}
\end{figure}

To analyze patterns, we compute two order parameters defined from spatial averages:
the average alignment of polarity $p_1=\langle \cos 2 \phi_\alpha \rangle$
and the average magnetization $p_2=\langle \rho_\alpha \cos 2 \phi_\alpha \rangle$.
Note that $-F p_2$ is equal to the average internal energy per site (see Eq.~\ref{eqEn1}). 
In Fig.~\ref{figEO}, we study how these order parameters change as a function of noise strength $\zeta$, a proxy for the effective temperature $T$.
Because $\zeta$ is proportional to $\sqrt{T}$ while the diffusion coefficient $D$ is proportional to $T$,
we vary parameters so that $D=d\zeta^2$, the constant $d$ being a function of mobility and dissipative coefficient.
In the absence of transport of polarity molecules ($\xi_\rho=0$), the two order parameters simultaneously decrease with noise strength, similar to classical models. 
However, they split when transport is allowed ($\xi_\rho>0$).
Average alignment, $p_1$, shows a drop when noise becomes positive, while magnetization, $p_2$, persists over a finite range of noise strengths. 
This split reflects the formation of spatial patterns, which have low global alignment and high global magnetisation. 
It occurs at finite values of noise, that we call intermediate.
 At very high noise, patterns disappear.
Fig.~\ref{figEO} also shows that increasing transport coefficient $\xi_\rho$ and decreasing diffusion $d$ sustain patterns and delay their disappearance.
\begin{figure}
\centerline{\includegraphics[width=4cm,angle=0]{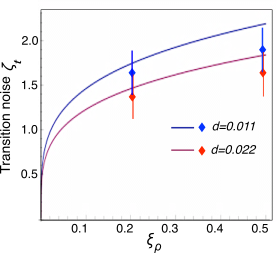}}
\caption{Noise at transition to non-uniform patterns in Fig.~\ref{figEO}, shown as a function of transport constant $\xi_\rho$: adiabatic approximation (lines) and numerical solutions (diamonds) with error bars representing uncertainties. }
\label{figTr}
\end{figure}

The transition between uniform state and patterns can be investigated analytically in the adiabatic limit where $\phi$ varies in time much faster than $\rho$.
$\phi$ then follows the Boltzmann distribution associated with the energy $2\xi_a F \rho_\alpha \cos 2 \phi_\alpha$. 
The time average, $m$, of $\cos 2\phi_\alpha$ is therefore given by $m=I_1(e)/I_0(e)$, where $I_0$ and $I_1$ are modified Bessel functions and $e=F \rho_\alpha/k_B T$. 
$m$ vanishes for $\rho_\alpha=0$, meaning that the orientation is temporally random, and $m=1$ for large $\rho$, meaning that the orientation follows the external field, as observed in Fig.~\ref{figE1}B,D.
We expand the equations around a homogenous state where the distribution of polarity magnitude is uniform.
The dynamics of magnitude $\rho$ is then described by a diffusion equation $\mathrm{d} \rho / \mathrm{d}t = (D - c\xi_\rho F^2 / k_B T) \Delta \rho$,
where $c=1/2$  for $e \ll 1$.
The homogenous state is stable only if $D>c \xi_\rho F^2 / k_BT$ or equivalently $\zeta > \zeta_t=(2 \xi_a c F^2 \xi_\rho / d)^{1/4}$,
which is in semi-quantitative agreement with numerical solutions (Fig.~\ref{figTr}).

{\it Patterns driven by the interaction between neighbors. }
\begin{figure}
\centerline{\includegraphics[width=7.7cm,angle=0]{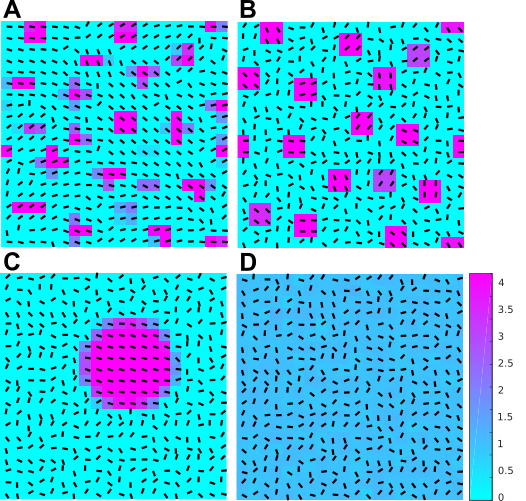}}
\caption{
Steady state configurations of polarity fields when neighbors are coupled. The lattice size is $20\times20$. The polarity orientation at each site is shown by a black bar. The polarity magnitude is color-coded with the same scale for all panels. 
The constant of driven polarity transport $\xi_\rho=0.067$. Noise strength, $\zeta$, varies between panels: (A) 0, (B) 0.73, (C) 1.46, (D) 1.95. The diffusion constant is proportional to the square of noise strength: $D=d\zeta^2$, with $d=0.11$. 
 }
\label{figPSN1}
\end{figure}
We now remove the external field to focus on the role of coupling between neighboring cells (although many of the observations hold with an external field). 
The dynamic equation for polarity orientation becomes
\bea
\frac{d\phi_\alpha}{dt}&=& 2K\xi_a \rho_\alpha \sum_{\beta}\rho_\beta \sin 2(\phi_\beta-\phi_\alpha) +\eta_\alpha (t)\textrm{,}
\label{eqD2}
\eea
while its magnitude follows Eq.~(\ref{eqD1}) with $f$ replaced by
\bea
 f_{\alpha}=K\sum_{\beta} \rho_\beta \cos 2(\phi_\alpha-\phi_\beta)\textrm{,}
\label{eqD3}
\eea
with summation over all neighbors of cell $\alpha$.

When transport of polarity molecules is allowed, the behavior of our model differs from the classical XY model.
Fig.~\ref{figPSN1} shows typical steady-state configurations of polarity fields.
Starting from a random distribution of polarity, the polarity magnitude becomes localized in small regions. 
Like in the first part of the paper, polarity orientation is aligned in high magnitude regions and appears random elsewhere.
However, polarity orientation varies from domain to domain, with no apparent correlation. 
Alignment appears stronger when $\xi_\rho$ is smaller.
Domains merge into bigger domains when noise strength is increased (Fig.~\ref{figPSN1}C). 
Domains cease to exist above a critical noise strength and the system reaches a configuration with roughly uniform polarity magnitude and random orientation (Fig.~\ref{figPSN1}D).

\begin{figure}
\centerline{\includegraphics[width=7.7cm,angle=0]{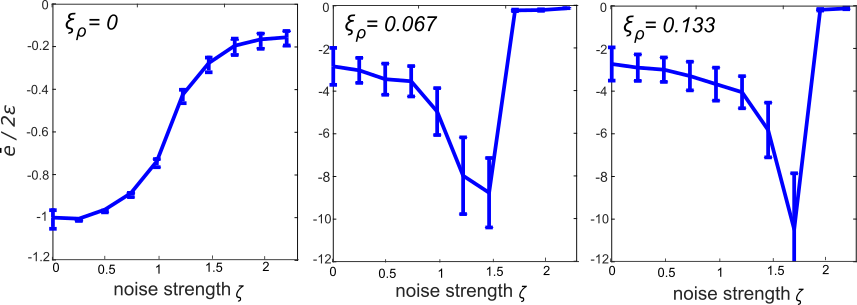}}
\caption{Average energy per site $\bar e=E/N^2$, normalized by $2\epsilon=2K \bar \rho^2$, in the steady state as a function of noise strength. The diffusion constant is proportional to the square of noise strength: $D=d\zeta^2$, with $d=0.11$. 
}
\label{figOPN}
\end{figure}

\begin{figure}
\centerline{\includegraphics[width=8cm,angle=0]{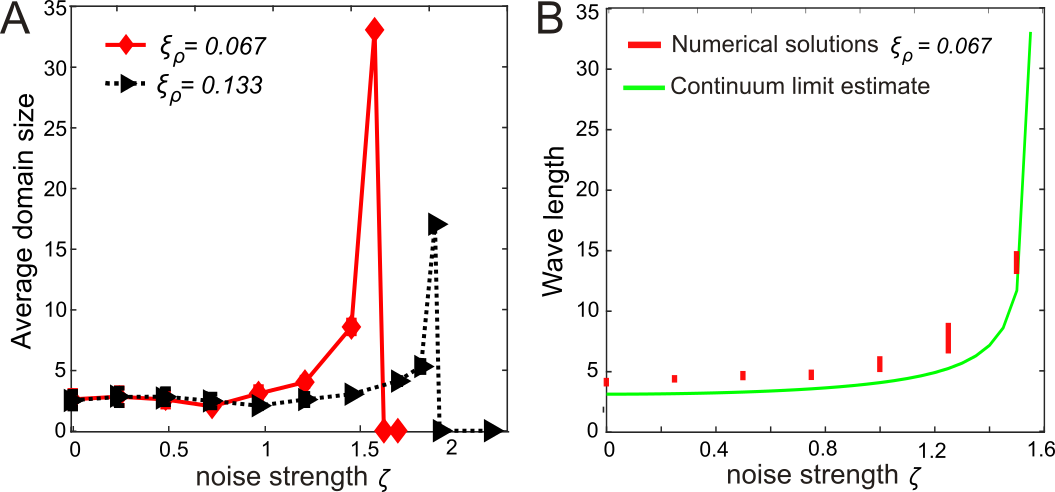}}
\caption{Size of steady-state patterns as a function of noise strength for a  $100\times100$ lattice. 
(A) Average diameter of domains with high density for two values of the dissipation coefficient $\xi_\rho$ in numerical solutions.
Error bars (SD) are smaller than markers.
(B) Comparison of the dominant wavelength between numerical solutions (shown as a range of the most robust wavelengths) and the continuum limit estimate.
$D=d\zeta^2$, with $d=0.11$ for all.}
\label{figRsize}
\end{figure}

We now analyze system behavior quantitatively. 
Fig.~\ref{figOPN} illustrates the steady state internal energy (which differs from the free energy) as a function of noise strength for different values of the driven transport constant, $\xi_\rho$. 
As expected, when $\xi_\rho=0$, the average internal energy increases with increasing noise as in classical models. 
For $\xi_\rho>0$, the average energy slowly decreases with noise strength, then drops sharply before rising to its maximum value corresponding to a network with a homogenous distribution of polarity magnitudes and random phase. 
The size of high magnetization domains hardly changes for intermediate noise strength and increases rapidly near the transition.  (Fig.~\ref{figRsize}A, obtained with bigger lattice of $100\times100$).
The wavelength behaves similarly to domain size, it remains finite even when noise strength vanishes and it increases at the transition (Fig.~\ref{figRsize}B, see Supplementary Note for details). 

To better understand these patterns, we consider the continuum limit of the model.
The energy can be approximated as
$E = K/2 \int \mathrm{d}x\mathrm{d}y  \{-4\rho^2/a^2 + (\nabla \rho)^2+ 4\rho^2 (\nabla\phi)^2\}$, where $a$ is lattice size.
The dynamics of polarity magnitude, linearized around a uniform state, follows
$d\rho/dt= (D- 4 K\xi_\rho)/a^2 \Delta \rho - K \xi_\rho \Delta^2 \rho$. The uniform state is unstable whenever
$4 K \xi_\rho>D$ and the most unstable wavelength is $\pi a/({1-\frac{D}{4K\xi_\rho}})^{1/2}$. 
This accounts for the disappearance of patterns at high noise, while predicted wavelengths agree with numerical solutions (Fig.~\ref{figRsize}B). 

%

{\it Conclusion. }
Our results demonstrate how adding intercellular transport of polarity magnitude to the XY model qualitatively affects its dynamics and results in the formation of patterns,
consisting of confined regions with high magnitudes of polarity, within a finite range of effective temperatures,
reminiscent of transitions observed in polymer suspensions~\cite{PhysRevLett.97.090602,PhysRevLett.101.218303,B923942A}. 

Future extensions could account for a non-conserved polarity field.
 It would be interesting to investigate system behavior with source and sink terms in the dynamic equations. 
To be closer to a biological tissue, we could consider lattices with different topologies, such as hexagonal, although we do not expect qualitative changes because the continuous limit of the model is well-defined as long as interactions are ferromagnetic ($K>0$).

Our model may apply to specific biological systems or to other polar media. 
Although there is no evidence of direct transport of polarity proteins in biological tissues, other molecules may be transported that influence polarity magnitude.
In this context, our study theorizes a novel mechanism for the self-organization of long distance patterns, that may be relevant to animal skin appendages or to hairs in plant epidermis. 
This mechanism might serve as a conceptual framework to think about such experimental systems, or as a starting point to develop more realistic models, for instance with additional mobile molecules that may mediate cell-cell coupling.
Finally, since our model is based on the general XY model that has been used extensively to describe systems such as magnetic rotors or spin glasses~\cite{RevModPhys.58.801,PhysRevLett.43.1754}, our extension could be significant for some of these systems.

This work was supported by the Agence National de la Recherche through the MorphoLeaf  [project$\#$ ANR-10-BLAN-1614]  and WallMime [project$\#$ ANR-17-CE20-0023-02] grants.
We gratefully acknowledge C. Godin and V. Mirabet for support and discussions, V. Zaburdaev, \' E. Rold\'an, and G. Ingram, for reviewing or proofing the manuscript prior to submission, and anonymous reviewers for helping to significantly improve the manuscript.

\clearpage
\bibliographystyle{apsrev4-1}
 \InputIfFileExists{foo.bbl}
 
\clearpage


\onecolumngrid
\section{Supplementary Note}

 \paragraph{ Numerical methods. }
We numerically solve the equations on an $N \times N$ square network with lattice size $a$, considering periodic boundary conditions. We use Euler integration method for stochastic systems. 
The initial conditions are such that the orientation of polarity at each site is a random variable with a uniform distribution in the interval $(0, \pi)$. 
Two cases are considered for the initial magnitude of polarity. 
In the main case, polarity magnitude is a random variable with a uniform distribution. 
In a particular case, polarity magnitude is localized in a square of four sites in the center. 
The spatial average of polarity magnitude is the same for the two initial conditions.

In our numerical analysis, we do not calculate the Laplacian directly. 
Instead, to ensure that polarity magnitudes remain positive, we approximate the flux from site $ \alpha$ to its neighbor $\beta$ by
\bea
 J_{ \alpha, \beta}= -\frac{\xi_\rho}{a} (f_ \alpha - f_ \beta) +\frac{D}{a}(\rho_ \alpha- \rho_ \beta).
 \eea 
 In such a square lattice, each site can also be defined by two indices $(i,j)$ representing the position along $x$ and $y$ axes. 
 The neighbors of site $\alpha$ are then indexed by $(i+1,j)$, $(i-1,j)$, $(i,j+1)$, and $(i,j-1)$.
For practical reasons, we assume that a site has zero outgoing flux if its polarity magnitude is smaller than a threshold, which we generally take as $1\%$ of the average polarity magnitude, though the value does not affect the results as far as it is small.

 We consider the diffusion constant $D=\mu k_B T$, where $\mu$ is the mobility and $T$ is an effective temperature accounting for random noise in the system. This effective temperature can be greater than thermal energy. 
The noise $\eta_\alpha (t)$ is an uncorrelated random noise  with zero mean and a white noise spectrum $\langle \eta_\alpha (t) \eta_\beta (t')\rangle= 2 \xi_a k_B T \delta_{\alpha,\beta} \delta(t-t')$. 
In our numerics, at each step $\eta$ is computed from an independent random variable with Gaussian normal distribution of zero mean and standard deviation of $\zeta/\sqrt{dt}$. 
Here $dt$ represents the time step  and $\zeta=\sqrt{2\xi_a k_B T}$ is proportional to square root of an effective temperature. 
We consider the diffusion constant to be proportional to the square of noise strength $D=d \zeta ^2$, where $d$ is a constant. 
We integrate the dynamical equation of polarity magnitude and angles using a time step of $dt=10^{-4}$.  

For our numerical solutions and the presented plots, we use the following normalization of parameters.
Lattice size $a$  and time unit scale $1/(2 \xi_a \epsilon)$ are used correspondingly to normalize length and time. Here $\epsilon$ is the scaling energy, given by $F \bar \rho$ and $K \bar \rho^2$, respectively, in the first and second cases. $\bar \rho$ represents the average polarity magnitude per site. 
Moreover, $\xi_\rho$  is normalized by $2  \xi_a  a^2  \bar \rho^2$,
$\zeta$  is normalized by $\sqrt{2  \xi_a \epsilon}$,
and $d$ is normalized by $a^2  $.

\begin{figure}
\centerline{\includegraphics[width=16.cm,angle=0]{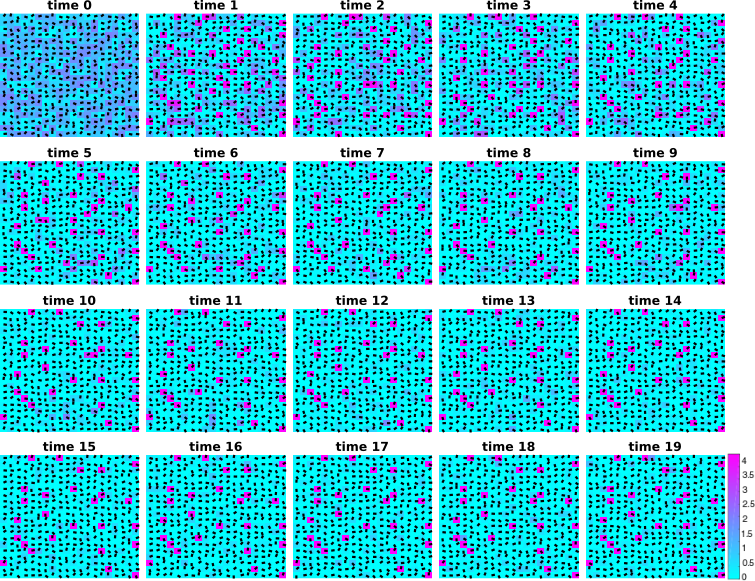}}
\caption{An example of a polarity field dynamics in a lattice of size $20\times20$  in the presence of an external field, initialized by a {\it random} distribution of polarity magnitude. Each time frame represents $4*10^4$ time steps. The polarity orientation at each site is shown by a black bar. Magnitude of polarity is color-coded with the same scale for all panels. It corresponds to the data presented in Fig.~1D in the main text.
 }
\label{figDy1}
\end{figure}
\begin{figure}
\centerline{\includegraphics[width=16.cm,angle=0]{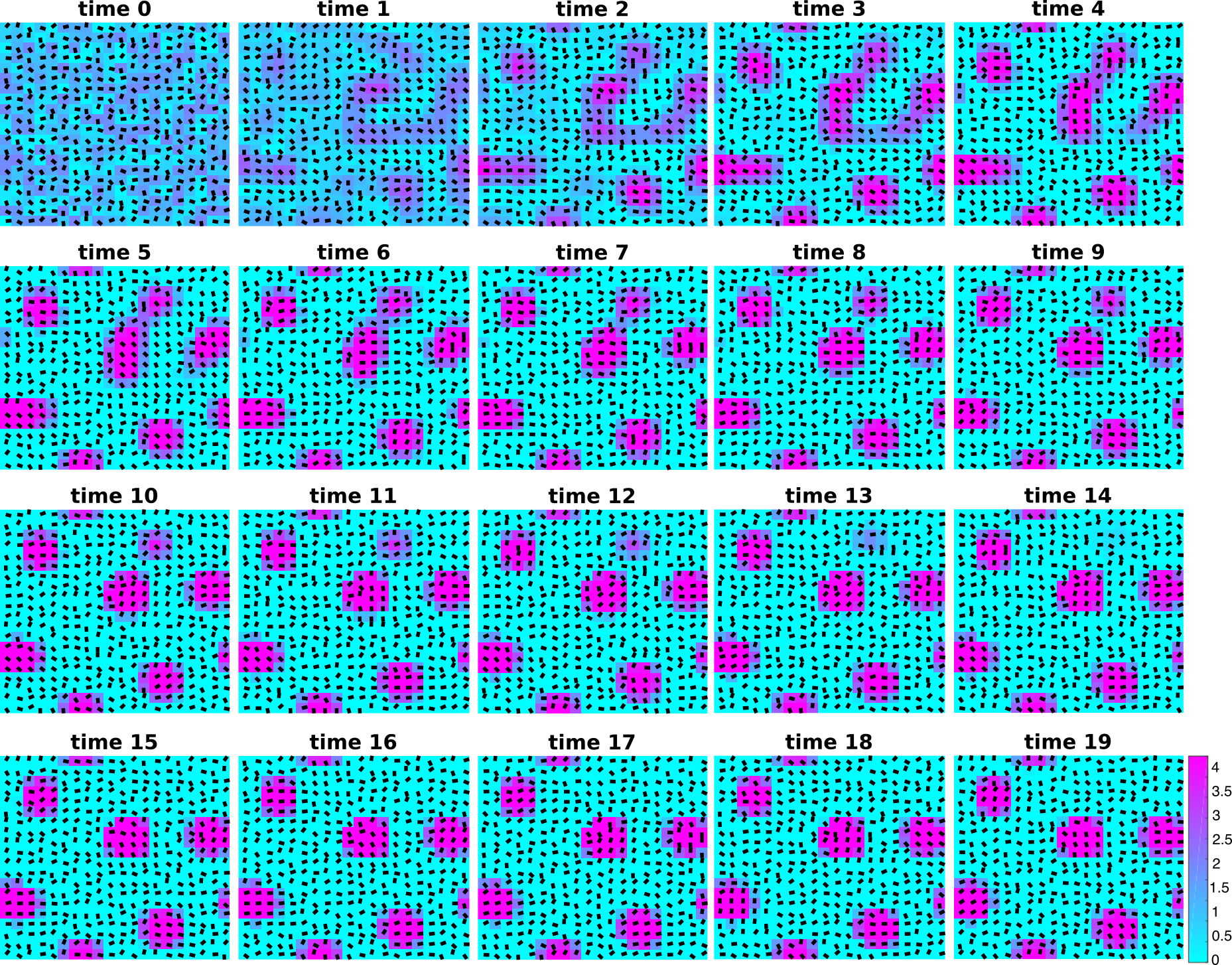}}
\caption{An example of a polarity field dynamics in a lattice of size $20\times20$ with coupling between neighbors  initialized by a {\it random} distribution of polarity magnitude. Each time frame represents $10^5$ time steps. The polarity orientation at each site is shown by a black bar. Magnitude of polarity is color-coded with the same scale for all panels. 
$d=0.11$ and $\xi_\rho=0.067$. $\zeta=1.22$. } 
\label{figDy}
\end{figure}
\begin{figure}
\centerline{\includegraphics[width=10.cm,angle=0]{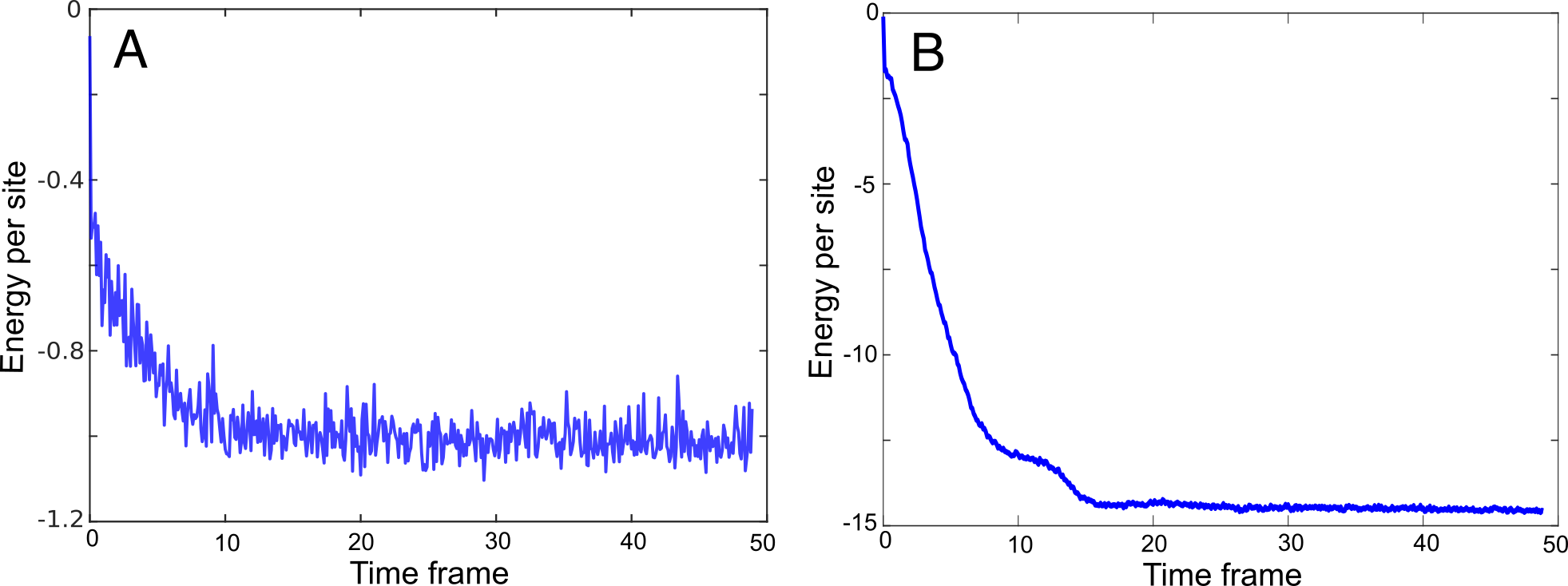}}
\caption{Time evolution of average energy per site for the lattice shown in (A) Fig.~\ref{figDy1} and (B) Fig.~\ref{figDy}.}
\label{figEnDy}
\end{figure}

\paragraph{ Transient behavior and steady state. } 
In this study, we are mostly interested in the steady-state behavior of the system. 
 Therefore, we solve the equations until a steady configuration is reached. 
 As an illustration, we show here typical transient behavior of the system, starting from a random configuration.
 Fig.~\ref{figDy1} shows the dynamics of the network in the presence of external field, when there can be exchange of polarity magnitude between neighbors. 
 Fig.~\ref{figDy} represents network dynamics for the second case with coupling between neighbors in the absence of external field. 
 This figure reveals how domains are formed and become stable during time driven by the exchange between neighbors. 
Following the transient, polarity magnitude is relatively fixed and exchanges between neighbors hardly occur. 
Polarity orientation may still vary, most notably in regions with low polarity magnitude and at high noise. 
However, we are always careful that the state variables, such as order parameters and energy, have reached a steady state. 
Fig.~\ref{figEnDy} shows that the energy of the same systems decays and indeed reaches a plateau.

 \paragraph{Estimate of transition between uniform and localized states.}  
We analyze transition between uniform states with homogenous distribution of polarity magnitude and non-uniform sates in which polarity magnitude is localized in some sites.
In the first case, in the presence of external field, we estimate the transition point where the order parameters $p_1$ and $p_2$ reunite. 
 This happens at finite noise level for non-zero values of  the polarity magnitude transport constant $\xi_\rho>0$ (see Fig. 2 in the main text). 
 The interval of noise strength in our numerics is then considered as an estimate of the maximum uncertainty of the transition noise.

 \paragraph{ Initial condition.}  
\begin{figure}
\centerline{\includegraphics[width=7.5cm,angle=0]{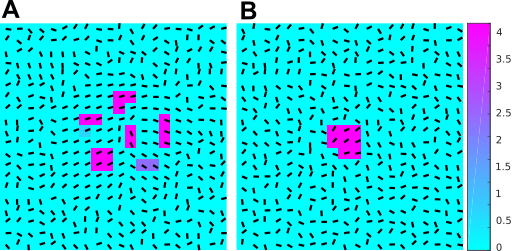}}
\caption{Steady state configurations of polarity fields with coupling between neighbors for the initial condition of polarity magnitude {\it localized} in the center. The lattice size is $20\times20$. The polarity orientation at each site is shown by a black bar. The polarity magnitude is color-coded with the same scale for all panels. 
The constant of driven polarity transport $\xi_\rho=0.067$. Noise strength, $\zeta$, varies between panels: 0 (A), 0.73 (B). The diffusion constant is proportional to the square of noise strength: $D=d\zeta^2$, with $d=0.11$. 
}
\label{figIniC}
\end{figure}
We note that,  steady state solutions can depend on initial configuration when noise is small. 
As an illustration, in the second case with neighbor coupling we start from a configuration where the polarity molecules are localized in the center.
Figs.~\ref{figIniC}A-B show that the polarity molecules spread out even in the absence of diffusion. 
However, the final configuration appears to remain pinned near the center, as compared to Fig.4A-B in the main text.
Therefore, at low noise, the final state of the system may be influenced by pre-patterns, i.e. by the patterns present in the initial condition.  
 \paragraph{ Analysis of domain size.} 
 We use a simple method to quantify the average size of patterns. 
 First, we calculate the diameter of domains as a measure of their size.
 We consider a threshold of polarity magnitude to define border of domains. 
 This threshold is unimportant because domain size hardly varies with the value of the threshold within a reasonable range. 
 In our numerics, we chose to use a threshold of $3$. 
 For each domain, the diameter is estimated as the largest border-to-border distance along the $x$ axis. 
 This is a legitimate estimate since domains are relatively symmetric. 
 In each lattice the total number of domains is denoted by $N_c$. 
 We provide the density of domains $N_c/N^2$, where $N$ is lattice size (Fig.~\ref{figRDEn}).
 We calculate the mean and standard deviation of the size of domains in each lattice (Fig. 6 in the main text), that is relatively very small. 

\begin{figure}
\centerline{\includegraphics[width=6.cm,angle=0]{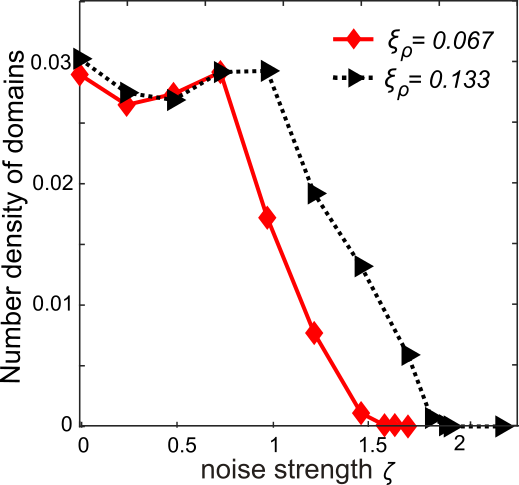}}
\caption{Number density of domains as a function of noise strength for a lattice of size $100\times100$ in steady-state limit. The data corresponds to the solutions presented in Fig.~6A in the main text.
}
\label{figRDEn}
\end{figure}

\begin{figure}
\centerline{\includegraphics[width=7.cm,angle=0]{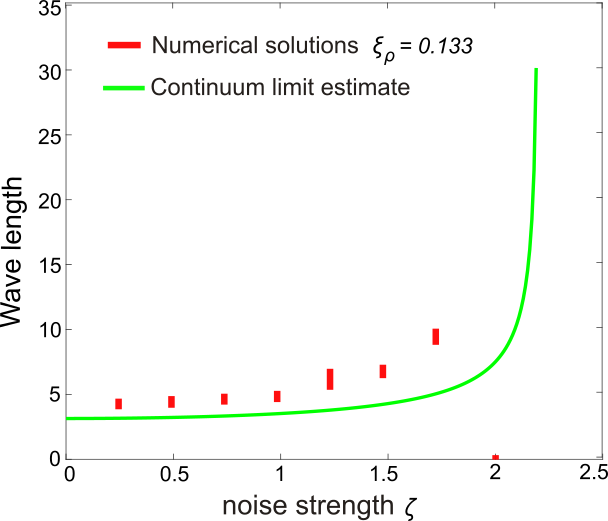}}
\caption{Comparison of the dominant wavelength between numerical solutions and the continuum limit estimate $\xi_\rho=0.133$.
Numerical solutions correspond to the steady state solution for a lattice of size $100\times100$ as presented in Fig.~6A in the main text.  }
\label{figWL2}
\end{figure}

Moreover, we analyze the patterns by computing the Fourier transform of the polarity density. 
For a network of size $N\times N$ with periodic boundary conditions, the Fourier coefficients of $\rho$ are given by
%
%
%
\bea
 \tilde \rho(n_x,n_y)=\sum_{i, j} \rho(i,j) e^{2\pi \mathrm{i}  (i n_x+jn_y)/N}.
\label{eqDphi}
\eea
Here $ \tilde \rho(n_x,n_y)$ is a complex number for any set of integers $n_x,n_y$. 
We define $n=\sqrt{n_{x}^{2}+n_{y}^{2}}$ and find $n^*$ that maximizes $| \tilde \rho(n)|$. 
The most robust wavelength is then given by $N/n$.
Because of the uncertainty in finding the position of the maximum of $| \tilde \rho(n)|$ in our numerical solutions, we show a range for the most robust wavelength (Fig.~\ref{figWL2} and Fig.~6B in the main text).

\end{document}